\documentclass[aps,pra,showpacs]{revtex4}
\usepackage{epsfig}

\newcommand{\beq}{\begin{equation}}
\newcommand{\eeq}{\end{equation}}
\newcommand{\beqa}{\begin{eqnarray}}
\newcommand{\eeqa}{\end{eqnarray}}
\newcommand{\nin}{\noindent}
\newcommand{\calP}{{\cal{P}}}

\begin{document}

\title{Quasilinearization approach to quantum mechanics}

\author{R.~Krivec$^1$
       and
       V.~B.\ Mandelzweig$^2$
       }

\affiliation{$^1$J. Stefan Institute, P.O.\ Box 3000, 1001 Ljubljana, Slovenia\\
         $^2$Racah Institute of Physics, Hebrew University, Jerusalem 91904, 
	 Israel}

\begin{abstract}
\widetext
\smallskip
The quasilinearization method (QLM) of solving nonlinear differential
equations is applied to the quantum mechanics by casting the Schr\"{o}dinger
equation in the nonlinear Riccati form.  The method, whose mathematical
basis in physics was discussed recently by one of the present authors (VBM),
approaches the solution of a nonlinear differential equation by
approximating the nonlinear terms by a sequence of the linear ones, and is
not based on the existence of some kind of a small parameter. It is shown
that the quasilinearization method gives excellent results when applied to
computation of ground and excited bound state energies and wave functions
for a variety of the potentials in quantum mechanics most of which are not
treatable with the help of the perturbation theory or the $1/N$ expansion
scheme. The convergence of the QLM expansion of both energies and wave
functions for all states is very fast and already the first few iterations
yield extremely precise results. The precison of the wave function is
typically only one digit inferior to that of the energy. In addition it is
verified that the QLM approximations, unlike the asymptotic series in the
perturbation theory and the $1/N$ expansions are not divergent at higher
orders.
\end{abstract}

\pacs{02.30.Mv, 04.25.Nx, 11.15.Tk}

\maketitle


\section{Introduction}

Realistic physical calculations usually are impossible without different
approximation techniques. Correspondingly expansions in small parameters,
statistical, variational and majority of numerical methods belong to the
arsenal of the modern physics.

Since many equations of physics are either nonlinear or could be cast in the
nonlinear form, the possibility of adding to this arsenal an additional very
powerful approximation technique applicable to nonlinear problems was
pointed out in a series of recent papers \cite{VBM,KM,MT}. It is called the
quasilinearization method (QLM) and its iterations are constructed to yield
rapid quadratic convergence and often monotonicity.  The quasilinearization
method was developed many years ago in theory of linear programming by
Bellman and Kalaba \cite{K59,BK65} as a generalization of the Newton-Raphson
method \cite{CB,RR} to solve the systems of nonlinear ordinary and partial
differential equations. Its modern developments and examples of applications
to different fields of science and engineering are given in a recent
monograph \cite{LV98}.

In the original works of Bellman and Kalaba \cite{K59,BK65}, however, the
convergence of the method has been proven only under rather restrictive
conditions of small intervals and bounded, nonsingular forces \cite{VBM}
which generally are not fulfilled in physical applications. This could
explain an extremely sparse use of the technique in physics, where only a
few examples of the references to it could be found
\cite{C67,AIC96,J88,RV87,HR83}. Recently, however, it was shown \cite{VBM}
by one of the present authors (VBM) that a different proof of the
convergence can be provided which allows to extend the applicability of the
method to realistic forces defined on infinite intervals with possible
singularities at certain points. This proof was generalized and elaborated
in the subsequent works \cite{KM,MT}.

In the first paper of the series \cite{VBM}, the analytic results of the
quasilinearization approach were applied to the nonlinear Calogero equation
\cite{C67} for the scattering length in the variable phase approach to
quantum mechanics, and the results were compared with those of the
perturbation theory and with the exact solutions.  It was shown that the
$n$-th QLM approximation sums exactly $2^{n}-1$ terms of the perturbation
theory while a similar number of terms are summed approximately. The number
of the exactly reproduced perturbation terms thus doubles with each
subsequent QLM approximation, which, of course, is a direct consequence of a
quadratic convergence.

The numerical calculation of higher QLM approximations to solutions of the
Calogero equation with different singular and nonsingular, attractive and
repulsive potentials performed in the next work \cite{KM} has shown that
already the first few iterations provide accurate and numerically stable
answers for any values of the coupling constant and that the number of
iterations necessary to reach a given precision increases only slowly with
the coupling strength. It was verified that the method provides accurate and
stable answers even for super singular potentials for which each term of the
perturbation theory diverges and the perturbation expansion consequently
does not exist.

In the third paper of the series \cite{MT} the quasilinearization method was
applied to other well known typical nonlinear ordinary differential
equations in physics, such as the Blasius, Duffing, Lane-Emden and
Thomas-Fermi equations which have been and still are extensively studied in
the literature. These equations, unlike the nonlinear Calogero equation for
the scattering length \cite{C67} considered in references \cite{VBM,KM},
contain not only quadratic nonlinear terms but various other forms of
nonlinearity and not only the first, but also higher derivatives. It was
shown that again just a small number of the QLM iterations yield fast
convergent and uniformly excellent and stable numerical results.

The goal of the present work is to apply the quasilinearization method to
quantum mechanics by casting the Schr\"{o}dinger equation in the nonlinear
Riccati form and calculating the QLM approximations to bound state energies
and wave functions for a variety of potentials, most of which are not
treatable with the help of the perturbation theory or the $1/N$ expansion
scheme. We show that the convergence of the QLM expansion for both energies
and wave functions is very fast and that already the first few iterations
yield extremely precise results. In addition it is verified that the higher
QLM approximations, unlike those in $1/N$ expansion method, are not
divergent at any order.

The paper is arranged as follows: in the second chapter we present the main
features of the quasilinearization approach to the solution of the
Schr\"{o}dinger equation, while in the third chapter we consider the
application of the method to computations for the Coulomb, Hulthen,
P\"{o}schl-Teller, logarithmic, double-well, anharmonic oscillator, linear
and different power potentials such as $r^{3/2}$ and $r^5$. The results and
their comparison with other calculations, convergence patterns, numerical
stability, advantages of the method and its possible future applications are
discussed in the final, forth chapter.

\section{Quasilinearization approach to the solution of the 
Schr\"{o}dinger equation}

The quasilinearization method (QLM) solves a nonlinear $n$-th order ordinary
or partial differential equation in $N$ dimensions as a limit of a sequence
of linear differential equations. The idea and advantage of the method
 is  based on the fact that linear equations can often be
solved analytically or numerically using superposition principle while there
are no useful techniques for obtaining the general solution of a nonlinear
equation in terms of a finite set of particular solutions.

The main feartures and equations of the method, appropriate for physics
applications, are summed in Refs.\ \cite{VBM,KM,MT}. In this paper we will
follow these references since the derivation there is not based, unlike the
derivation in Refs.\ \cite{K59,BK65}, on the smallness of the interval and
on the  boundedness  of the nonlinear term and its
functional derivatives, the conditions which usually are not fulfilled in
physics.

We would like to use the method in quantum mechanical calculations with the
central potential $V(r)$. In order to do this we have to rewrite the
corresponding radial Schr\"{o}dinger equation


\beq
-{{\hbar^2}\over{2m}} \chi''(r) +
\biggl[ V(r) + {{l(l+1)\hbar^2}\over{2mr^2}} \biggr]\chi(r) = E \chi(r)
\label{eq:Scheq}
\eeq
in nonlinear form. Here $\chi(r)=rR(r)$ and $R$ is the radial part of the
wave function. Setting $\hbar=1$, $\kappa^2=2m|E|$, and
$U(r)=2mV(r)+{l(l+1)}/{r^2}$, we obtain the bound state and scattering
Schr\"{o}dinger equations

\beq
\frac{d^2\chi(r)}{dr^2} - (\kappa^2+U(r))\chi(r)=0, \ E < 0
\label{eq:Schnegeq}
\eeq
and

\beq
\frac{d^2\chi(r)}{dr^2} + (\kappa^2-U(r))\chi(r)=0, \ E > 0
\label{eq:Schposeq}
\eeq
with the boundary conditions at the origin

\beq
\chi(r)\mathop{\sim}_{r\rightarrow 0} r^{l+1} 
\label{eq:bca}
\eeq
\nin

and at the infinity for the potentials falling off at large  $r$

\beqa
\chi(r)&&\mathop{\sim}_{r\rightarrow \infty} e^{-\kappa r}, 
\ E < 0,
\label{eq:bcbneg} \\
\chi(r)&&\mathop{\sim}_{r\rightarrow \infty}
\sin(\kappa r-\frac{\pi l}{2}+\delta_l), \ E> 0. 
\label{eq:bcbpos}
\eeqa

For potentials behaving at large  $r$  as 
$\lambda^2 \ln {r}/{R}$  or $\lambda^2 r^p$ with positive $R, p$
and $\lambda$ the boundary conditions at infinity should be changed
respectively to

\beq
\chi(r)\mathop{\sim}_{r\rightarrow \infty}
e^{-\lambda \int^r \sqrt{\ln{{r}/{R}}} dr} 
\label{eq:bcca}
\eeq
or
\beq
\chi(r)\mathop{\sim}_{r\rightarrow \infty}
e^{-\frac{2\lambda}{p+2}r^{p/2+1}}. 
\label{eq:bccb}
\eeq

The boundary condition (\ref{eq:bccb}) with $r$ changed to $|r|$ holds at both
boundaries $r=\pm \infty$ in the one dimensional problem for the double-well
potential ${{(r^2 - 16)^2}/{128}}$ considered among others in Ref.\
\cite{BB2000} where we look for both ground (symmetric) and first excited
(antisymmetric) solutions.

It is easy to check that the inverse dimensionless logarithmic derivative
$\phi$ of the wave function,

\beq
\phi(x)=\kappa \frac{\chi(r)}{\chi'(r)},
\label{eq:phi}
\eeq
satisfies the nonlinear Riccati equations

\beq
\frac{d\phi(x)}{dx}=1-(1+W(x)) \phi^2(x), \ E < 0
\label{eq:Ricnegeq}
\eeq
and

\beq
\frac{d\phi(x)}{dx}=1+(1-W(x)) \phi^2(x), \ E > 0,
\label{eq:Ricposeq}
\eeq
\nin
where $x=\kappa r$ and $W(x)=U(x/\kappa)/\kappa^2$ are the dimensionless
variable and potential, respectively.

To avoid poles of $\phi(x)$ at the bound state energies it is convenient to
define (see \cite{VBM,KM} and the references therein) a new function $u(x)$ with the help of the equation

\beq
\phi(x)=-\tan u(x).  
\label{eq:tan}
\eeq
The corresponding equations for $u(x)$ have the forms
\beq
\frac{du(x)}{dx}=-1+(2+W(x)) \sin^2 u(x), \ E < 0
\label{eq:Ricnegu}
\eeq
and

\beq
\frac{du(x)}{dx}=-1+ W(x)\sin^2 u(x), \ E > 0.  
\label{eq:Ricposu}
\eeq
Similar  types  of equations were derived earlier by
Drukarev \cite{Dru}, Bergmann \cite{Ber}, Olsson \cite{Ols}, Kynch
\cite{Kyn}, Franchetti \cite{Fra}, Spruch \cite{Spr}, Dashen \cite{Das}, 
Calogero \cite{C67} and Babikov \cite{B67}.

The boundary conditions for the function $u(x)$, in view of  Eqs.\
(\ref{eq:bca}-\ref{eq:bcbpos}),  respectively, reduce to

\beqa
u(x)&&\mathop{\sim}_{x\rightarrow 0} -\frac{x}{l+1}, 
\label{eq:phibca} \\
u(x)&&\mathop{\sim}_{x\rightarrow \infty}\frac{\pi}{4} - n\pi,\ E < 0
\label{eq:phibcb}
\eeqa
and
 
\beq
u(x) + x\mathop{\sim}_{x\rightarrow \infty}\frac{\pi l}{2} - \delta_l, E > 0. 
\label{eq:phibcbb}
\eeq
The boundary conditions (\ref{eq:bcca}) and (\ref{eq:bccb}) which for
the inverse logarithmic derivative $\phi(x)$ have the forms

\beq
\phi(x)\mathop{\sim}_{x\rightarrow\infty}-\frac{\kappa}{\lambda\sqrt{\ln{\frac{x}{\kappa R}}}}
\rightarrow 0,
\label{eq:phibcca}
\eeq
and
\beq
\phi(x)\mathop{\sim}_{x\rightarrow \infty} -\frac{x^{-{p}/{2}}}{\lambda}
\rightarrow 0,
\label{eq:phibccb}
\eeq
respectively, for the function $u(x)$ therefore read
\beq
u(x)\mathop{\sim}_{x\rightarrow \infty}-n\pi .
\label{eq:ubcb}
\eeq
Here and in Eq.\ (\ref{eq:phibcb}) $n$ obviously denotes the number of the
excited state with $n=1$ corresponding to the ground state, $n=2$ to the
first excited state etc. The minus sign in front of $n$ follows from the
fact that in the regions of $r$ where $V(r)< E$ in view of Eqs.\
(\ref{eq:Ricnegu}) and (\ref{eq:Ricposu}) the derivative ${du}/{dx}$ is
negative and $u(x)$ is decreasing. Since its value at the origin is zero
$u(x)$ stays negative which determines the sign in front of $n$.

Returning to the variable $r$ and defining a new function $a(r)$ which has
the dimension of length with the help of the relation
$\phi(x)=\kappa(r+a(r))$ and substituting it into Eqs.\ (\ref{eq:Ricnegeq})
and (\ref{eq:Ricposeq}) we obtain the equations

\beq
\frac{da(r)}{dr}=-(\kappa^2+U(r))(r+a(r))^2, \ E<0   
\label{eq:Ricreqneg}
\eeq
and
\beq
\frac{da(r)}{dr}=(\kappa^2-U(r))(r+a(r))^2, \ E> 0   
\label{eq:Ricreqpos}
\eeq
which are very similar to the Calogero equation

\beq
\frac{da(r)}{dr}=-2 m V(r) \ (r+a(r))^2, 
\label{eq:Caleq}
\eeq
where $a(r)$ has the meaning of the variable $s$-wave scattering length
\cite{C67}. These equations are obviously a generalization of the Calogero
Eq.\ (\ref{eq:Caleq}) for arbitrary values of $l$ and $\kappa$ and reduce to
it when $l$ and $\kappa$ are equal to zero.

The QLM prescription \cite{VBM,KM,MT,K59,BK65} determines the $k+1$-th
iterative approximation $u_{k+1}(x)$ to the solution of the first order
nonlinear equation in one variable

\beq
\frac{du(x)}{dx}=f(u(x),x), \ u(0)=0 
\label{eq:uxf}
\eeq
as a solution of the linear equation

\beqa
u'_{k+1}(x)&=&f(u_k,x) + (u_{k+1}(x)-u_k(x))f_u(u_k,x), 
\label{eq:difreq} \nonumber
\\
u_{k+1}(0)&=&0,
\label{eq:difreq1} 
\eeqa
where the functional $f_u(u,x)={\partial{f(u,x)}}/{\partial{u}}$ is a
functional derivative of the functional $f(u(x),x)$.

The analytical solution of this equation is

\beqa
u_{k+1}(x)=\int_{0}^{x}ds(f(u_k&&(s),s)-f_u(u_k(s),s)u_k(s)) 
\label{eq:iqlma} \nonumber
\\
&&\times \exp\int_{s}^{x}dtf_u(u_k(t),t).   
\label{eq:iqlma1} 
\eeqa
The sequence $u_k(x),\ k=0,1,2,...$ of QLM iterations satisfying Eqs.\
(\ref{eq:difreq}) and (\ref{eq:iqlma}), converges {\it uniformly and
quadratically} to the solution $u(x)$ of Eq.\ (\ref{eq:uxf}) if the initial
guess for the zeroth iteration is sufficiently good. In addition, for
strictly convex (concave) functionals $f(u(x),x)$ the difference
$u_{k+1}(x)-u_k(x)$ is strictly positive (negative) which establishes the
{\it monotonicity} of the convergence from below (above), respectively. The
exact conditions of the convergence and the monotonicity for the realistic
physical conditions of forces defined on infinite intervals with possible
singularities at certain points are formulated in Ref.\ \cite{VBM}. One can
also prove \cite{C67} that in the quasilinear approximation the energy in
the Schr\"{o}dinger equation satisfies the Rayleigh-Ritz variational
principle which ensures the quadratic convergence in the QLM energy
computations.

We will limit ourselves here to the bound state calculations with Eqs.\
(\ref{eq:Ricnegu}) for the negative energy bound states and
(\ref{eq:Ricposu}) for positive energy bound states which are somewhat more
complicated than scattering calculations since in the former case the
boundary condition at infinity determines a discrete spectrum.

For the negative energies, Eq.\ (\ref{eq:Ricnegu}), the functionals 
$f(u(x),x)$,
$F(u(x),x) \equiv f_u(u(x),x)$ and
$G(u(x),x){\equiv}f(u(x),x)-u(x)f_u(u(x),x)$ are given by

\beqa
f(u(x),x)&=&-1+(2+W(x)) \sin^2 u(x), 
\label{eq:uxfneg} \\
F(u(x),x)&=&(2+W(x)) \sin 2 u(x),  
\label{eq:uxfder} 
\eeqa
and

\beqa
G(u(x),x)=&&-1+(2+W(x))\ \sin u(x) 
\label{eq:uxfg} 
\nonumber \\
&&\times [\sin u(x)-2 u(x) \cos u(x)],  
\label{eq:uxfg1}
\eeqa
so that Eqs.\ (\ref{eq:difreq}) and (\ref{eq:iqlma}) respectively have the
forms

\beq
u'_{k+1}(x)-u_{k+1}(x) F(u_k(x),x) = G(u_k(x),x),
\label{eq:uxfred}
\eeq
and

\beq
u_{k+1}(x)=\int_{0}^{x}dsG(u_k(s),s)\exp\int_{s}^{x}dtF(u_k(t),t).   
\label{eq:uxfrei}
\eeq

For the positive energies the same equations (\ref{eq:uxfneg})-(\ref{eq:uxfg})
hold with $(2+W(x))$ replaced everywere by $W(x)$.

\section{QLM bound state calculations and their comparison with the
$1/N$ expansion method and exact solutions}

In the previous chapter we have cast the Schr\"{o}dinger equation in the
nonlinear Riccati form and wrote the linear equations and the boundary
conditions appropriate for the bound state calculations with
the quasilinearization method.

In this chapter we consider examples of different singular and nonsingular
attractive interactions which, in view of their large coupling constants,
are not treatable with the help of the perturbation theory and for most of 
which the $1/N$ expansion series are asymptotically divergent as has been
shown in Ref.\ \cite{BB2000}.

Namely, we apply the quasilinearization method to computations with the
Coulomb, Hulthen, P\"{o}schl-Teller, logarithmic, anharmonic oscillator,
linear and different other power potentials such as $r^{3/2}$ and $r^5$ as
well to the one-dimensinal double-well potential ${{(r^2-16)^2}/{128}}$, and
we compare the wave functions and the bound state energies obtained by the
quasilinearization method (QLM) with their exact values and with results
obtained in the $1/N$ expansion theory. To show that the method works
equally well also for excited states we calculate in the Coulomb, linear and
double well potentials the first few excited states as well.

The calculations were done using the differential formulation, Eq.\
(\ref{eq:difreq}), of the QLM iteration, for the simple reason that the
adaptive numerical integration \cite{NAGlib} together with interpolation
proved faster than the integral formulation (\ref{eq:iqlma}), mainly due to
the processor time taken by the evaluation of the exponential in Eq.\
(\ref{eq:iqlma}). For each QLM iteration number $k$, $k=0,1,2,3,\ldots k_m$,
numerical integration was performed from $x=0$ to the matching point $x=x_m$
and from the upper bound $x=x_{\rm{U}}$ to $x=x_m$.

Let us denote the set of iteration-integration parameters by
${\calP}=\{k_m,x_{\rm{U}},N_i,\ldots\}$, where $k_m$ is the maximum QLM
iteration index, $x_{\rm{U}}$ is the upper bound of the interval, and $N_i$
is the number of interpolation points in each of the two subintervals
$(0,x_m)$ and $(x_m,x_{\rm{U}})$.

The computation was done in two steps. In the first step, $x_m$, the
starting values of parameters, ${\calP}_0$, and a $\kappa$ value,
$\kappa_0$, near the expected eigenvalue were prescribed. On the last QLM
iteration ($k=k_m$) the absolute difference between the left-hand side (LHS)
and right-hand side (RHS) solutions,
$D_{{\calP}_0}(\kappa_0;x_m)=|u_{k_m}^{{\calP}_0,{\rm{LHS}}}(x_m,\kappa_0)-
u_{k_m}^{{\calP}_0,{\rm{RHS}}}(x_m,\kappa_0)|$, was calculated. The whole
process was then performed with a new set of parameters ${\calP}_1$, where
$k_m$, $x_{\rm{U}}$, etc.\ were increased. This was repeated until some
number $M$ of steps, when $D_{{\calP}_M}(\kappa_0;x_m)$ was stabilized to a
required accuracy.

In the second step, the parameter set ${\calP}_M$ thus optimized was used to
find the zero of $D_{{\calP}_M}(\kappa;x_m)=0$ as a function of $\kappa$:
the QLM iteration was first performed for two $\kappa$ values lying on
opposite sides of the expected eigenvalue, and the QLM iteration
($k=1,2,3,\ldots $) was then repeated for each new $\kappa$ value until
$D_{{\calP}_M}(\kappa;x_m)=0$.

In this process the value of $x_m$ was kept constant, which had the
consequence that the RHS interval $(x_m,x_{\rm{U}})$ was increasing. Both
solutions tend to become unstable near $x=x_m$ on their respective sides, if
the respective interval is too large. It turned out however that it was
possible to leave $x_m$ unchanged, except that $x_m$ typically had to
increase with the number of the excited state. On the other hand, as is
evident from the Figures, the starting values of parameters (${\calP}_0$),
in particular $k_m$, had to be large enough to overcome the divergent
behavior of the solutions near $x=x_m$ already for the QLM iteration using
${\calP}_0$. It also turned out that the RHS solution quickly assumes the
correct value, thus allowing reasonably small $x_{\rm{U}}$, and actually
making the process rather independent of the exact value used for the
boundary condition at infinity.

The precision was controlled in the following way. The differential equation
solver \cite{NAGlib} was required to return $u(x)$ with the precision of the
order of $10^{-P_{\rm{ODE}}}$. The required precision of $D_{\calP}$ during
the optimization of $\calP$ was $10^{-P_{\calP}}$. $P_{\rm{ODE}}$ was taken
to be larger than $P_{\calP}$ by 1 to 3 to test stability.

\begin{figure}
\begin{center}
\epsfig{file=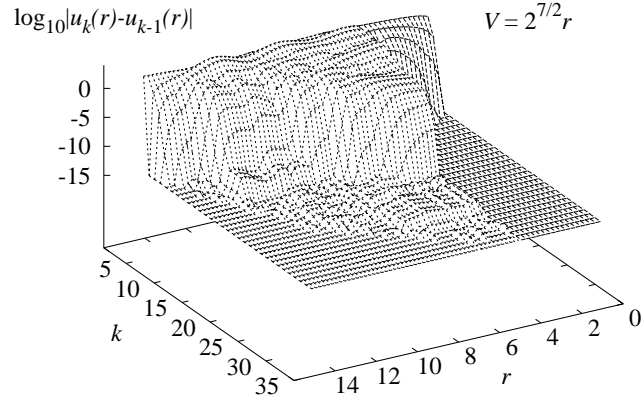,width=98mm}
\end{center}
\caption{
Convergence of the logarithm of the absolute value of the difference of two
successive QLM iterations $u_k(r)$ for all $r$ with the iteration index $k$
for the ground state of the linear potential $V=2^{7/2}r$, $m=1$. Here
$u(r)=\arctan(-\kappa\chi(r)/\chi'(r))$ and $\kappa=\sqrt{2mE}$. The
matching point is at $r=4$.}
\label{qlm2fig01}
\end{figure}
\begin{figure}
\begin{center}
\epsfig{file=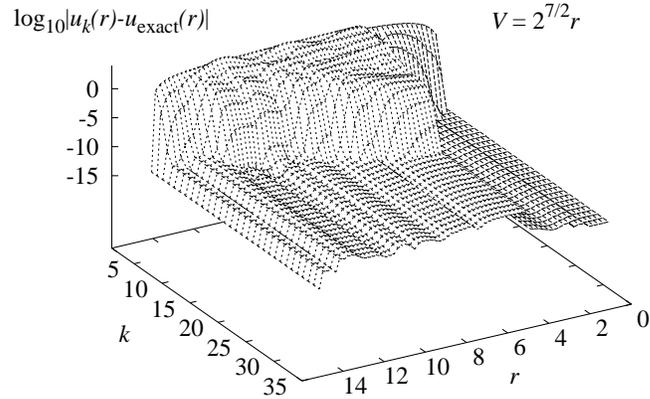,width=98mm}
\end{center}
\caption{
As in Fig.\ \ref{qlm2fig01}, but the convergence with respect
to the exact solution obtained by solving the differential equation
for $\phi$, Eq.\ (\ref{eq:Ricposeq}).}
\label{qlm2fig02}
\end{figure}
\begin{figure}
\begin{center}
\epsfig{file=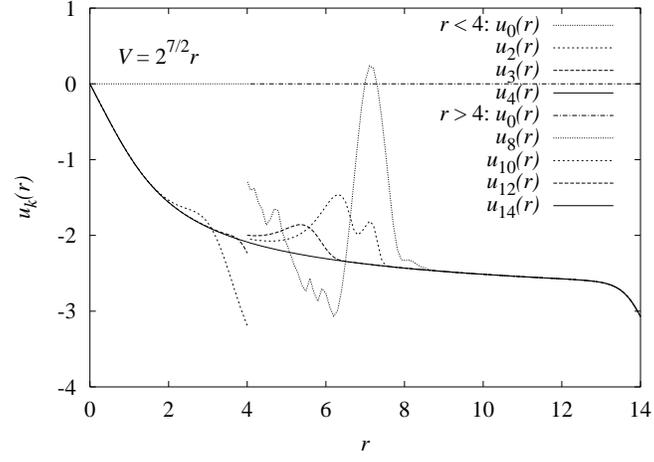,width=90mm}
\end{center}
\caption{
Convergence of the QLM iterations with the iteration index $k$ for the solution
of Fig.\ \ref{qlm2fig01}. Only a subset of iterations is presented
such that the highest ones are not distinguishable from the exact solution.}
\label{qlm2fig03}
\end{figure}
\begin{figure}
\begin{center}
\epsfig{file=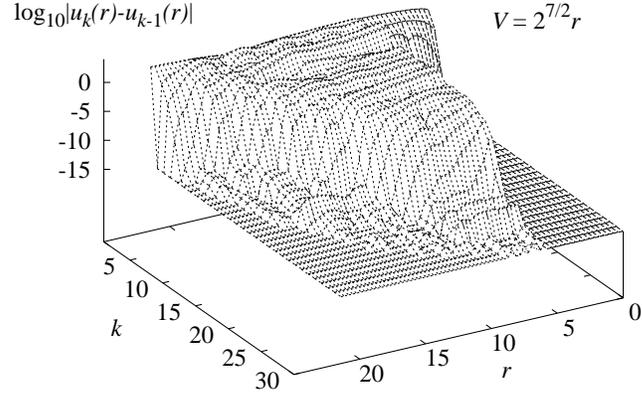,width=98mm}
\end{center}
\caption{
As in Fig.\ \ref{qlm2fig01}, but for the first excited state,
and the matching point being at $r=6$.}
\label{qlm2fig04}
\end{figure}
\begin{figure}
\begin{center}
\epsfig{file=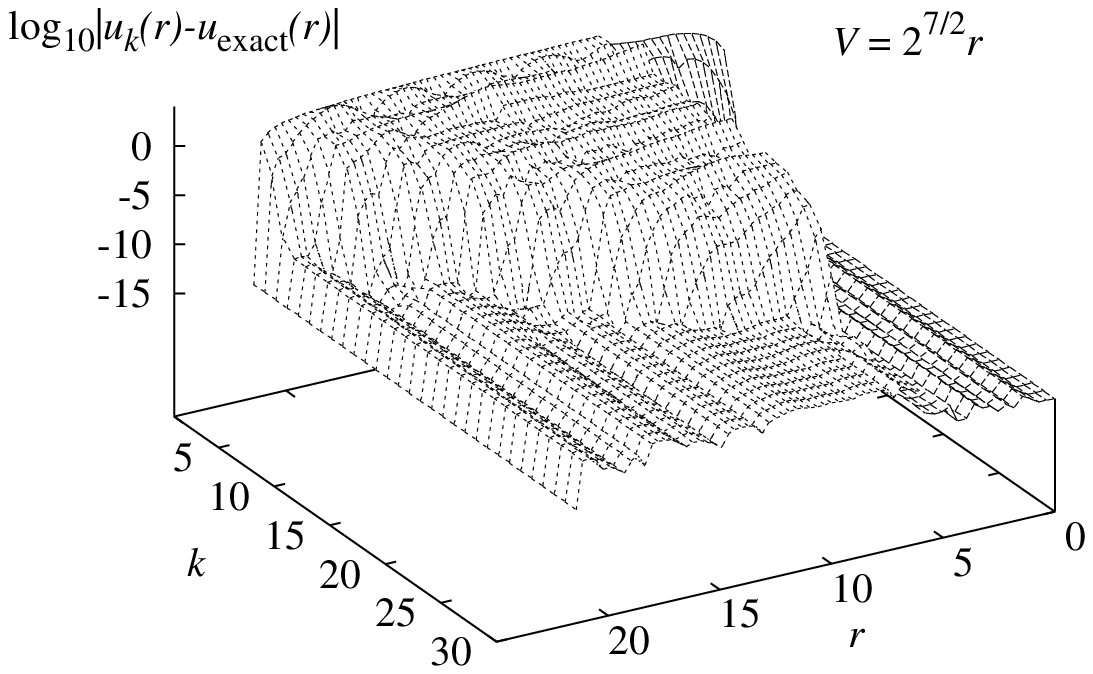,width=98mm}
\end{center}
\caption{
As in Fig.\ \ref{qlm2fig02}, but for the state of Fig.\ \ref{qlm2fig04}.}
\label{qlm2fig05}
\end{figure}
\begin{figure}
\begin{center}
\epsfig{file=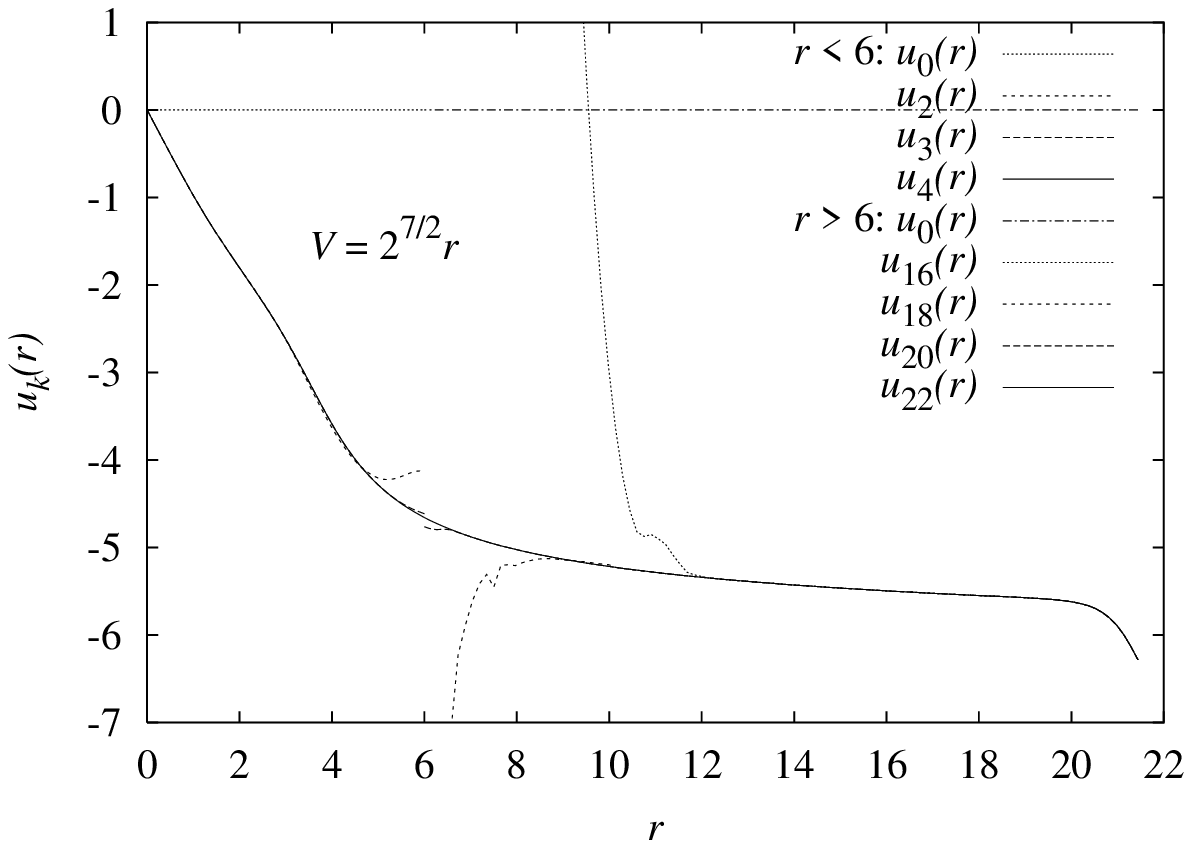,width=90mm}
\end{center}
\caption{
As in Fig.\ \ref{qlm2fig03}, but for the state of Fig.\ \ref{qlm2fig04}.}
\label{qlm2fig06}
\end{figure}
\begin{figure}
\begin{center}
\epsfig{file=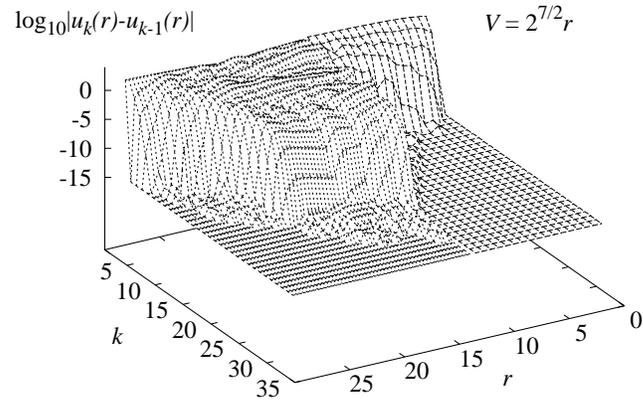,width=98mm}
\end{center}
\caption{
As in Fig.\ \ref{qlm2fig01}, but for the second excited state,
and the matching point being at $r=12$.}
\label{qlm2fig07}
\end{figure}
\begin{figure}
\begin{center}
\epsfig{file=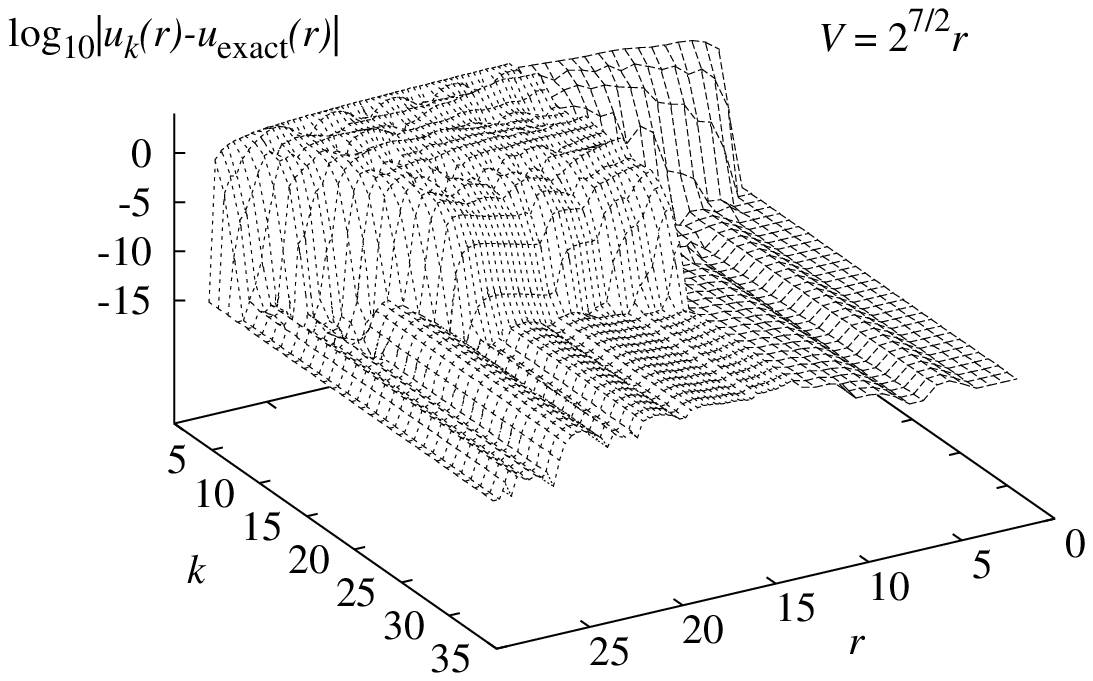,width=98mm}
\end{center}
\caption{
As in Fig.\ \ref{qlm2fig02}, but for the state of Fig.\ \ref{qlm2fig07}.}
\label{qlm2fig08}
\end{figure}
\begin{figure}
\begin{center}
\epsfig{file=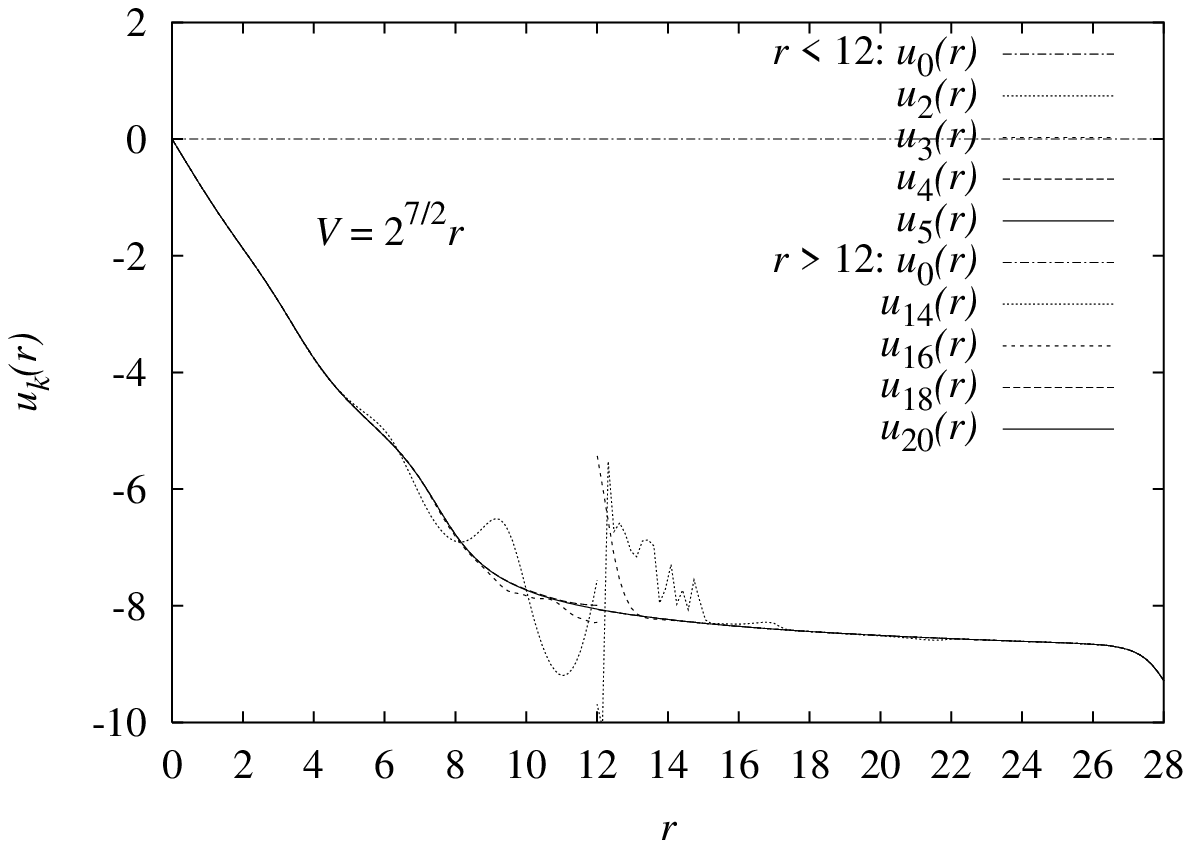,width=90mm}
\end{center}
\caption{
As in Fig.\ \ref{qlm2fig03}, but for the state of Fig.\ \ref{qlm2fig07}.}
\label{qlm2fig09}
\end{figure}
\begin{figure}
\begin{center}
\epsfig{file=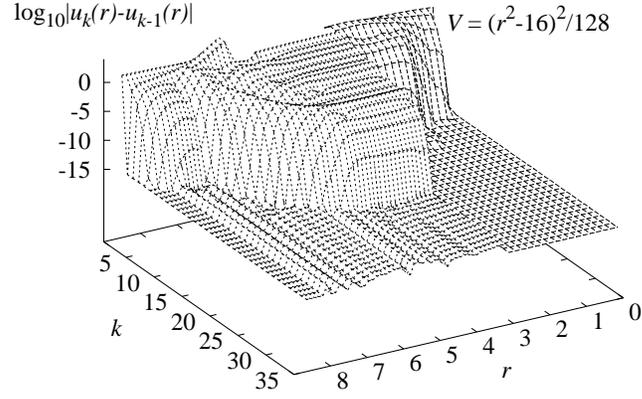,width=98mm}
\end{center}
\caption{
As in Fig.\ \ref{qlm2fig01}, but for the ground (symmetric) state in 
the double-well potential $V=(r^2-R^2)^2/(8R^2)$, $R=4$ and $m=1/2$.}
\label{qlm2fig10}
\end{figure}
\begin{figure}
\begin{center}
\epsfig{file=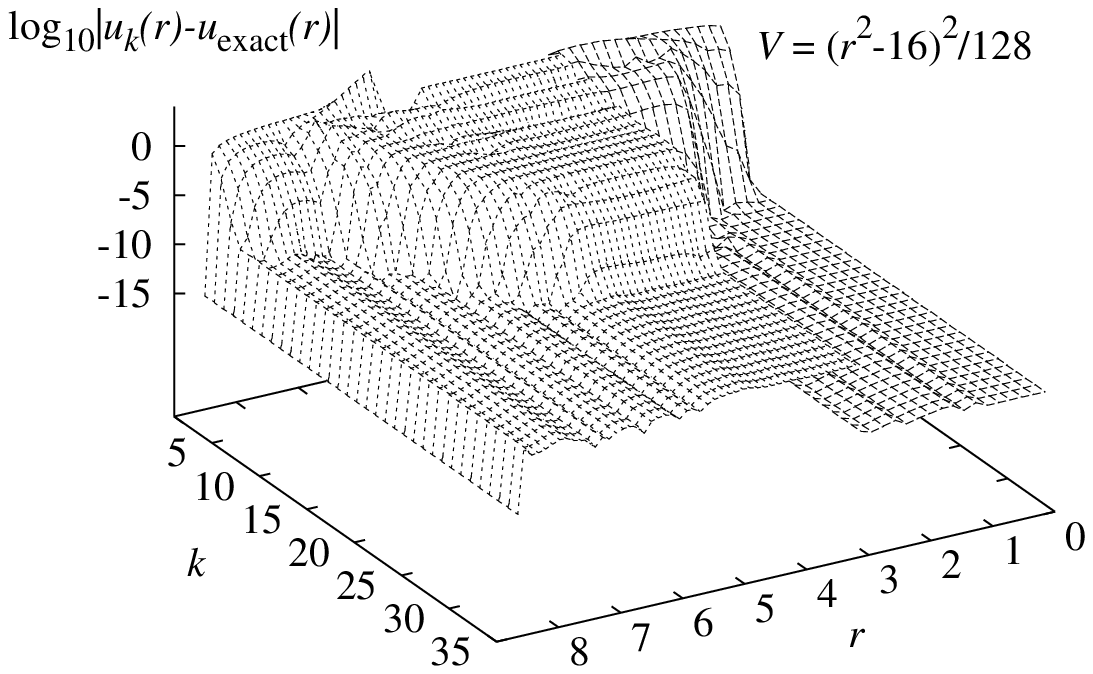,width=98mm}
\end{center}
\caption{
As in Fig.\ \ref{qlm2fig02}, but the potential and state of Fig.\ \ref{qlm2fig10}.}
\label{qlm2fig11}
\end{figure}
\begin{figure}
\begin{center}
\epsfig{file=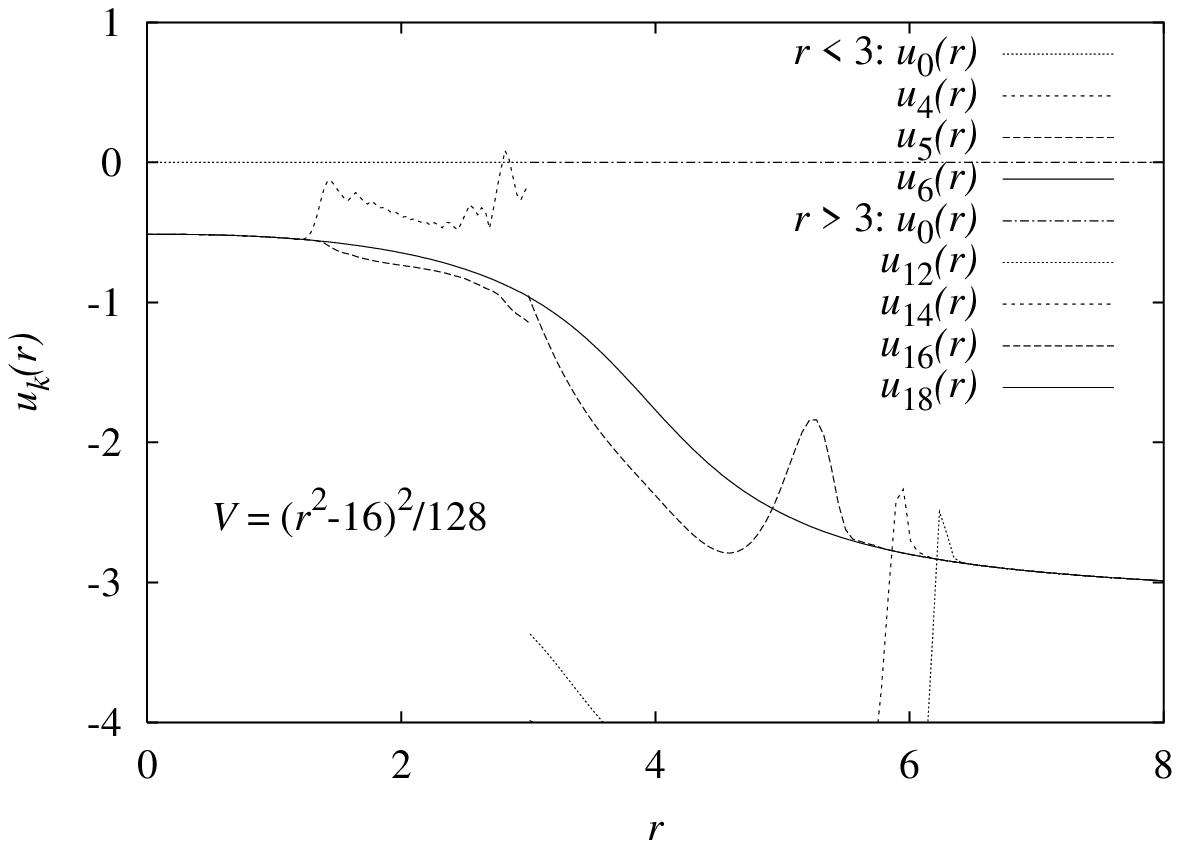,width=90mm}
\end{center}
\caption{
As in Fig.\ \ref{qlm2fig03}, but the potential and state of Fig.\ \ref{qlm2fig10}.}
\label{qlm2fig12}
\end{figure}
\begin{figure}
\begin{center}
\epsfig{file=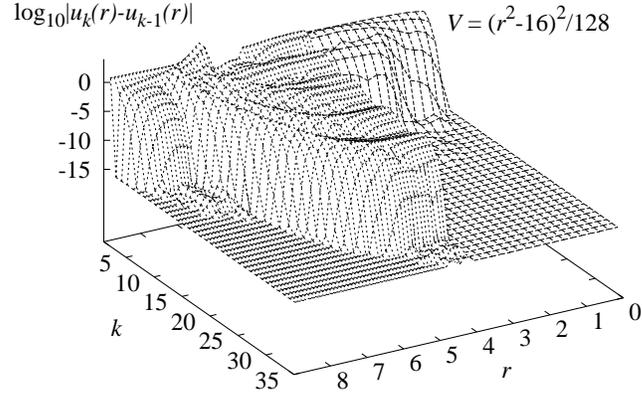,width=98mm}
\end{center}
\caption{
As in Fig.\ \ref{qlm2fig01}, but for the first excited (antisymmetric) state in 
the double-well potential $V=(r^2-R^2)^2/(8R^2)$, $R=4$ and $m=1/2$.}
\label{qlm2fig13}
\end{figure}
\begin{figure}
\begin{center}
\epsfig{file=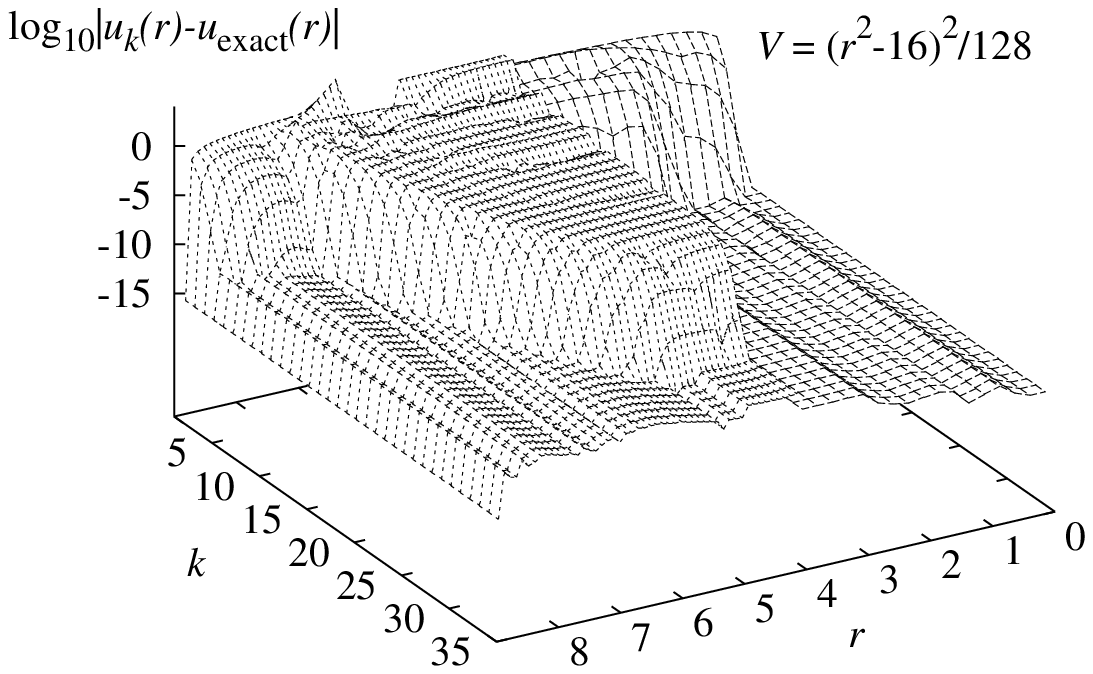,width=98mm}
\end{center}
\caption{
As in Fig.\ \ref{qlm2fig02}, but the potential and state of Fig.\ \ref{qlm2fig13}.}
\label{qlm2fig14}
\end{figure}
\begin{figure}
\begin{center}
\epsfig{file=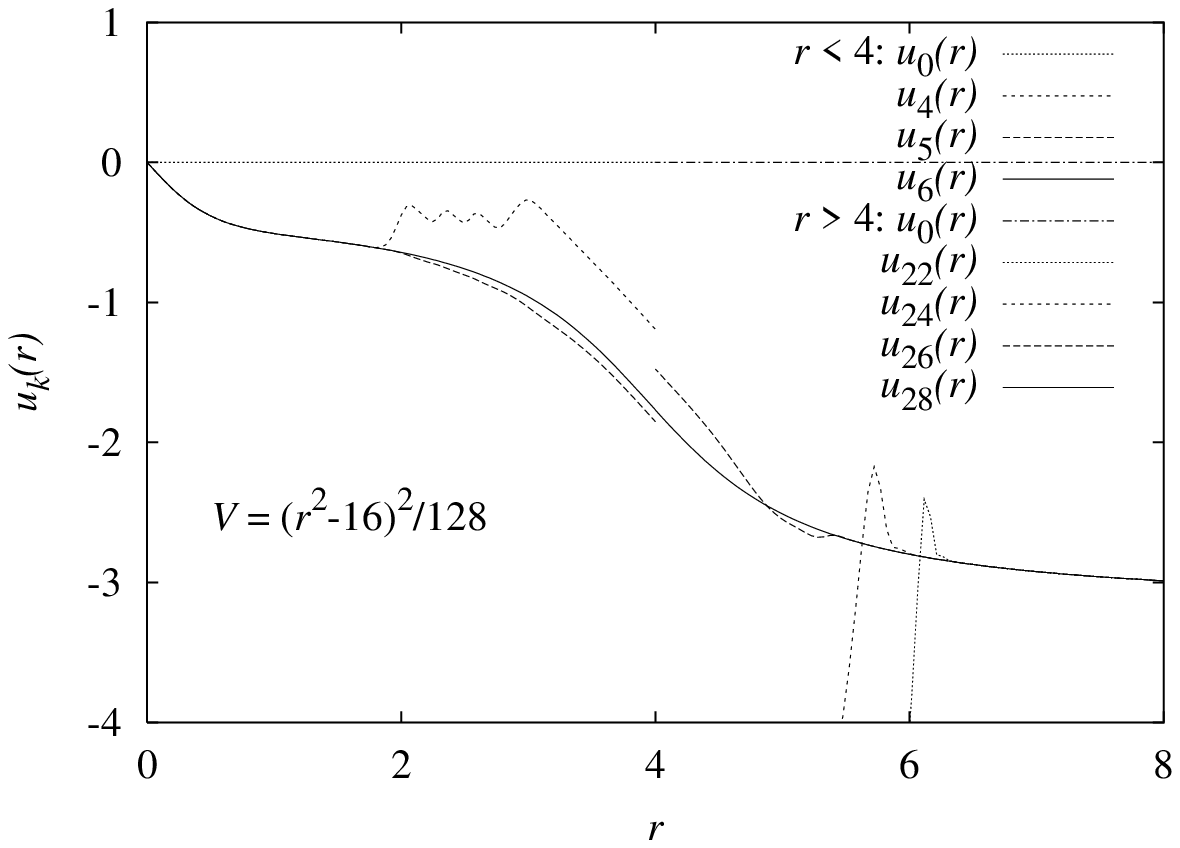,width=90mm}
\end{center}
\caption{
As in Fig.\ \ref{qlm2fig03}, but the potential and state of Fig.\ \ref{qlm2fig13}.}
\label{qlm2fig15}
\end{figure}

The results of the calculations are summarized in Table \ref{qlm2tab01} and
in Figs.\ \ref{qlm2fig01}-\ref{qlm2fig15}. The calculations are done for the
$s$-states since the calculations for $p$, $d$ states etc.\ have the same
degree of difficulty and could be performed in a similar fashion. In the
caption of the table $V(r)$ is the potential and $n$ denotes the number of
the excited state; $m$ is the mass of the particle and is given different
values for different potentials in order to enable comparison of the QLM
bound state energies with those obtained by the $1/N$ expansion method in
Ref.\ \cite{BB2000} where $m$ takes on values $m=1$ or $1/2$ depending on
the interaction. In the graphs of the convergence of $u_k(x)$ with iteration
index $k$ we present for clarity only those iterations which are
distinguishable from the final solution ($k=k_m$) on the graphs; the actual
number of iterations is higher in order to achieve greater wave function
precision. Figures which display the absolute differences between successive
iterations, $|u_k(r)-u_{k-1}(x)|$, or the differences between the successive
iterations and the exact solution, $|u_k(r)-u_{\rm exact}(x)|$, show the
results for the respective optimized parameters sets, $D_{{\calP}_M}$, and
for the last $\kappa$ value, {\it i.e.}, at the $E$ of the eigenvalue.

The required precision of $u(x)$, or the wave function, was $P_{\rm{ODE}}=9$
in all cases except in the logaritmic potential case where $P_{\rm{ODE}}=6$.
The number of digits in the values of $E$ in Table \ref{qlm2tab01} is the
number of stable digits when $P_{\calP}$ was increased up to
$P_{\rm{ODE}}+3$, except in the cases of Coulomb, P\"{o}schl-Teller and
H\"{u}lthen potentials, where we display an additional (the first incorrect)
digit.

From Table \ref{qlm2tab01} and Figs.\ \ref{qlm2fig01}-\ref{qlm2fig15} one
can conclude that QLM is extremely precise. Energies and the wave function
for both ground end excited states typically converge to the order of 10
significant digits after about $k_m=10 - 20$ iterations though the precision
of $E$ is about one digit more than the precision of the wave function. We
used the numbers of QLM iterations $k_m$ such that the precision of the
iteration itself, shown by the Figures displaying $|u_k(r)-u_{k-1}(x)|$, was
up to $10^{-15}$.

\section{Conclusion}

Our calculations confirm numerically the conclusion following from the proof
in Ref.\ \cite{MT} that once the quasilinear iteration sequence starts to
converge, it will continue to do so, unlike the perturbation expansions in
powers of the coupling constant or in powers of $1/N$, which are often given
by the asymptotic series and therefore converge only up to a certain order
and diverge thereafter. In particular, the $1/N$ expansions of the binding
energy of different ground and excited states given in Table
\ref{qlm2tab01}, are strongly divergent for logarithmic, double-well,
anharmonic oscillator, linear, $r^{3/2}$ and $r^5$ potentials at orders of
about 20 or higher or even before this as it was shown recently by
Bjerrum-Bohr \cite{BB2000}.

Based on our results of the QLM computations of the wave functions and bound
state energies for many different potentials, one can deduce the following
important features of the quasilinearization method in
the quantum mechanics:

\begin{itemize}

\item[(i)]

  The quasilinearization method solves the Schr\"{o}dinger equation by
  rewriting it in the nonlinear Riccati form and by approximating the
  nonlinear terms by a sequence of the linear ones. It is not based, unlike
  perturbation or $1/N$ expansion theories, on the existence of some kind of
  small parameter.

\item[(ii)]

  The quasilinearization method works equally well for both ground and
  excited states. It is extremely precise: binding energies and the wave
  functions converge to the order of 10 significant digits after about $10 -
  20$ iterations. Typically, the numerically obtained precison of the wave
  function is only one digit inferior to that of the energy.

\item[(iii)]
  
  Fast convergence of the QLM iterations to the exact solution confirms
  numerically the uniform and quadratic law of convergence proved in Refs.\
  \cite{VBM,KM,MT} for realistic physical interactions defined on infinite
  intervals with possible singularities at certain points of the intervals.

\item[(iv)]
  
  For convergence it is enough that an initial guess for the zeroth
  iteration is sufficiently good.  In all the examples considered in the
  paper the simplest initial guess of setting the logarithmic derivative of
  the wave function $\phi(x)$ equal to zero or to $x$ at the origin was
  enough to produce a rapid convergence.

\item[(v)] 

  By using the high numbers of QLM iterations $k_m$ such that the obtained
  iterative solution was extremely accurate, up to $10^{-15}$, it was
  numerically confirmed the statement earlier proved and verified in Ref.\
  \cite{MT} that once the quasilinear iteration sequence starts to converge,
  it will always continue to do so unlike the perturbation or $1/N$
  expansion series, which are asymptotically divergent. The
  quasilinearization method therefore always yields the required precision
  once a successful initial guess generates convergence after a few steps.

\end{itemize}

In view of all this, the quasilinearization method appears to be extremely
useful in quantum mechanics and in many cases more advantageous than the
perturbation theory or its different modifications, like expansion in
inverse powers of the coupling constant, the $1/N$ expansion, etc. Though in
this work only central potentials and one dimensional double well potential
were considered and thus only differential equations in one variable were
treated, the quailinearization method is able to solve the systems of
nonlinear ordinary and partial differential equations in $N$ variables and
could therefore be applicable to the solution of the Schr\"{o}dinger
equation with the noncentral potentials or to the $N$-body Schr\"{o}dinger
in $3N-3$ dimensions which will be subject of the future research.

\acknowledgments The research was supported by the
bilateral Cooperation Program of the Ministry of Science and Technology of
Slovenia (RK) and by the Israel Science Foundation grant 131/00 (VBM).

\pagebreak

\newpage

\begin{table}[hbt]
\begin{center}
\caption{%
QLM and exact binding energies E for different potentials.
$E$ are taken from citations in Ref.\ \protect\cite{BB2000}. $E_{p}$ and
$N_{p}$ are the energies by the $1/N$ perturbation method of Ref.\
\protect\cite{BB2000} and the corresponding ranges of $N$ where the $1/N$
expansion converges; a finite range means that the expansion diverges for
larger $N$; stable digits are given only. $n$ is the principal quantum
number of the state. The uncertainty in last digit is in brackets where
necessary for presentation. $m$ denotes the (reduced) mass of the particle.}
\begin{tabular}{|ccr|ll|lr|}
\hline
 $V$                  & $m$ &  $n$  & QLM          &     $E$       & $E_{p}$  & $N_{p}$ \\
\hline
$2^{7/2}r$            &  1  & 1 & ~9.352429642 & ~9.35243  & ~9.352   & 10--20 \\
                      &     & 2 & 16.35179778  & 16.3518   & 16.352   & 10--15 \\
                      &     & 3 & 22.08223931  & 22.08224  & 22(1)    & 10--28 \\
&&&&&&\\                                                             
$r^{3/2}$             & 1/2 & 1 & ~2.708092416 & ~2.70809  & ~2.71    & 14--15 \\
&&&&&&\\                                                             
$\ln r$               & 1/2 & 1 & ~1.044332    & ~1.0443   & ~1.04    & 113--14\\
&&&&&&\\
$-{{1}\over{r}}$      &  1  & 1 & ~0.499999999 & ~0.5      &$1\kern-3pt\pm\kern-3pt10^{-10}$ & 29--$\infty$ \\
                      &     & 2 & ~0.125000001 & ~0.125    &$0.25\kern-3pt\pm\kern-3pt10^{-8}$ & 29--$\infty$ \\

&&&&&&\\
$r^5$                 & 1/2 & 1 & ~4.089159315 & ~4.08916  & ~4.      & 6--7\kern4 pt\\
&&&&&&\\
$r^2+r^4$             & 1/2 & 1 & ~4.648812183 & ~4.64881  & ~4.6(2)  & 10--11\\
&&&&&&\\
${{(r^2 - 16)^2}\over{128}}$
                      &  1  & 1 & ~0.4830541244& 0.483053433     & ~0.48302 & 12--13\\
                      &     &   &              & 0.483053390     & & \\
&&&&&&\\
                      &  1  &  2  & ~0.4831482068&           &          & \\
&&&&&&\\
$-{{3}\over{\cosh^2 r}}$
                      &  1  & 1 & ~0.49999999998& ~0.5     &          & \\
&&&&&&\\
$-{{10}\over{\cosh^2 r}}$
                      &  1  & 1 & ~4.49999999991& ~4.5     &          & \\
&&&&&&\\
$-{{e^{-r/5}}\over{1-e^{-r/5}}}$
                      &  1  & 1 & 12.0050000001 & 12.005   &          & \\
&&&&&&\\
$-{{e^{-r}}\over{1-e^{-r}}}$
                      &  1  & 1 & ~0.125000000009 & 0.125  &          & \\
&&&&&&\\
\hline
\end{tabular}
\label{qlm2tab01}
\end{center}
\end{table}

\clearpage

\clearpage

\clearpage

\clearpage

\clearpage

\end{document}